\pdfoutput=1
\documentclass{report}

\usepackage[cmex10]{amsmath}
\usepackage{bbm}
\usepackage{graphicx}
\usepackage{latexsym}
\usepackage{comment}
\usepackage{bbm}
\usepackage{ mathrsfs }
\usepackage{ amssymb }
\usepackage{authblk}
\usepackage{epstopdf}
\usepackage[backend=biber]{biblatex}
\usepackage{url}

\usepackage{hyperref}

\usepackage{listings}
\usepackage{color}

\usepackage{algorithm}
\usepackage[noend]{algpseudocode}
\usepackage {tikz}
\usepackage{tkz-berge}
\usetikzlibrary {positioning}

\definecolor{dkgreen}{rgb}{0,0.6,0}
\definecolor{gray}{rgb}{0.5,0.5,0.5}
\definecolor{mauve}{rgb}{0.58,0,0.82}
\definecolor {processblue}{cmyk}{0.96,0,0,0}

\usepackage{xcolor}

\begin{document}

\title{Annual Interruption Rate as a KPI, its measurement and comparison}

%\author{\IEEEauthorblockN{Rohit Pandey}
%\IEEEauthorblockA{Azure Compute Insights\\
%Microsoft\\
%Redmond, WA\\
%ropandey@microsoft.com}
%\and
%\IEEEauthorblockN{Avnish Chhabra}
%\IEEEauthorblockA{Azure One Deploy\\
%Microsoft\\
%Redmond, WA\\
%avnishc@microsoft.com}
%\and
%\IEEEauthorblockN{Yingnong Dang}
%\IEEEauthorblockA{Azure Compute Insights\\
%Microsoft\\
%Redmond, WA\\
%yidang@microsoft.com}
%\and
%\IEEEauthorblockN{Abhishek Kumar}
%\IEEEauthorblockA{Azure OneDeploy\\
%Microsoft\\
%Redmond, WA\\
%abkmr@microsoft.com}
%\and
%\IEEEauthorblockN{Gil Lapid Shafriri}
%\IEEEauthorblockA{Azure Compute Insights\\
%Microsoft\\
%Redmond, WA\\
%gilsh@microsoft.com}
%\and
%\IEEEauthorblockN{Murali Chintalapati}
%\IEEEauthorblockA{Azure Compute Insights\\
%Microsoft\\
%Redmond, WA\\
%muralic@microsoft.com}
%}

\author{Rohit Pandey$^{*}$,
        Yingnong Dang$^{*}$,
        	Ali Vira$^{*}$,
        Aerin Kim$^{**}$,
        Gil Lapid Shafriri$^{*}$,
        Murali Chintalapati$^{*}$\\
$^{*}$Azure Compute Insights\\
$^{**}$BAG CTO Group\\
Email: \emph{\{ropandey,yidang,avira,ahkim,gilsh,muralic\}@microsoft.com}
}

\date{\today}

\maketitle

\tableofcontents{}
\pagebreak
\section*{Introduction}
Azure is Microsoft's cloud service, providing customers various computing resources like virtual machines (VMs), app services, website hosting capabilities, web jobs, storage and much more. The focus of Azure Compute is on virtual machines that customers can rent and use as their needs dictate, scaling up and down at will. The quality of service then is measured in terms of how available these machines are, how well they are performing, etc. Azure is dedicated to providing customers with the most highly available cloud and a sizable chunk of its employees are dedicated to this goal (with everyone else having it at the top of their minds).

So, Azure as an organization and all its employees want to make these virtual machines we rent to customers highly available. To make sure all these people are pulling in the same direction, they all need to agree on what exactly that means. Since there is an increasing trend for modern organizations to be data-driven (with Azure being no exception), it makes sense to express this goal in terms of numbers that can then be tracked across the organization, with everyone working towards moving them in the right direction. These numbers are called `Key Performance Indicators` or KPIs. 

This article is divided into two chapters. The first chapter describes failure rate as a KPI and studies its properties. The second one goes over ways to compare this KPI across two groups using the concepts of statistical hypothesis testing.

In section 1., we will motivate failure rate as a KPI (in Azure, it is dubbed `Annual Interruption Rate' or AIR. In section 3, we will discuss measuring failure rate from logs machines typically generate. In section 1.2, we will discuss the problem of measuring it from real-world data.

In section 2.1, we will discuss the general concepts of hypothesis testing. In section 2.2, we will go over some general count distributions for modeling Azure reboots. In section 2.3, we will go over some experiments on applying various hypothesis tests to simulated data. In section 2.4, we will discuss some applications of this work like using these statistical methods to catch regressions in failure rate and how long we need to let changes to the system `bake' before we are reasonably sure they didn't regress failure rate. 

\chapter{Annual Interruption Rate}

\section{Failure rate as a KPI}
Over the years, the core KPI used to track availability within Azure has shifted and evolved. For a long time, it was the total duration of customer VM downtime across the fleet. However, there were two issues with using this as a KPI:

\begin{itemize}
\item{It wasn't normalized, meaning that if we compare it across two groups with the first one having more activity, we can obviously expect more downtime duration as well.}
\item{It wasn't always aligned with customer experience. For example, a process causing many reboots each with a short duration wouldn't move the overall downtime duration by much. However, it is still a very poor experience for some customers whose workloads are sensitive to any interruptions. Customers running gaming workloads for example tend to fall into this category.}
\end{itemize}

Another KPI that would atleast address the second problem was count of customer VM interruptions. However, the issue remained that it wasn't normalized. As the size of the Azure fleet grows over time, we therefore expect the number of interruptions across the fleet to increase as well. But then, if we see the number of fleet-wide interruptions increasing over time, how do we tell how much of it is due to the size increasing and how much can be attributed to the platform potentially regressing?

To address these problems, a new KPI called the `Annual Interruption Rate' or AIR was devised. Before describing it, let's take a highly relevant detour into the concept of ``hazard rate" from Statistics. It can be interpreted as the instantaneous rate at which events are occurring, much like velocity is the instantaneous rate at which something is covering distance.

This rate can be expressed in terms of properties of the distribution representing times elapsing between the events of interest which in this case might be VM reboots. Since this time between two successive reboots is a random variable, we will express it as an upper-case, $T$. Since this notion of rates applies to any events, that is how we will refer to these `reboots'. If we denote the probability density function (PDF) of this random variable, $T$ by $f_T$ and the survival function (probability that the random variable, $T$ will exceed some value, $t$) by $S_T(t) = P(T>t)$, then the hazard rate is given by:

\begin{equation}h_T(t) = \frac{f_T(t)}{S_T(t)} \tag{1}\end{equation}

The way to interpret this quantity is that at any time $t$, the expected number of events the process will generate in the next small interval, $\delta t$ will be given by: $h_T(t) \delta t$. You can find a derivation of this expression in appendix A. Note again that this is an instantaneous rate, meaning it is a function of time. When we talk about the Azure KPI, we're not looking to estimate a function of time. Instead, given some interval of time (like the last week) and some collection of VMs, we want to get a single number. In reality, the rate will indeed probably vary from instant to instant within our time interval of interest. So, we want one estimate to represent this entire profile. 

It is helpful again to draw from our analogy with velocity. If a car were moving on a straight road with a velocity that is a function of time and we wanted to find a single number to represent its average speed, what would we do? We would take the total distance traveled and divide by the total time taken for the trip. Similarly, the average rate (let's denote it by $\lambda$) over a period of time will become the number of events we are modeling divided by the total time interval (say $\bigtriangleup t$). 

\begin{equation}\lambda = \frac{n}{\bigtriangleup t}\tag{2}\end{equation}

You can find a derivation of this 'average rate' in appendix A as well.

The `average rate' defined here is what the `AIR' (Annual Interruption Rate) KPI used within Azure is based on. It is the projected number of reboots/ other events (like blips and pauses, etc.) a customer will experience if they rent 100 VMs and run them for a year (or rent one VM and run it for 100 years; what matters is the VM-years). So, in equation (2), if we measure the number of VM-years for any scope (ex: entire Azure, a customer within Azure, a certain hardware, etc.). 

This definition in equation (2) is almost there, but is missing one subtlety related to VMs in Azure going down for certain intervals of time as opposed to being point-events. This means that the VM might be up and running for an interval of time and then go down and stay down for some other interval before switching back to up and so on. In the next section, we'll discuss ways to take this into account.

\section{Measuring failure rate}
In this section, we will consider how to actually calculate AIR (or failures rates) from Azure logs (or general machine logs). Let's see what these might look like.

\subsection{Machine data}

When considering machine failure, we need to measure it. That means there has to be some logging process that writes data about the functioning of the machine somewhere it can be aggregated.

This kind of data can generally fit into two categories, heartbeat data or state transition data (described below in detail). The former can easily be converted to the latter and the latter is easier to work with as well as more compressed. Generally, heartbeat data is the first thing the machine logs and these logs are then converted into state transition data for more sophisticated analysis.

\begin{itemize}
\item{\textbf{Heartbeat data}: A log from the machine sayings it's `alive' and either an explicit log saying it's `down' or no log at all for an extended period of time if things have gone so far south that it isn't even able to log. These logs are generally emitted at periodic time intervals. Like once every few seconds or minutes. Five minutes is quite typical.}
\item{\textbf{State Transition data}: This is a compressed version of the heartbeat data where we have one entry (or row) per \textit{interval} of the component being in one of the two states (UP or DOWN). So, if the component were DOWN for the whole day, instead of logging down heartbeats all day, there will be one entry (row in data table) saying the component was down for a duration of one day. See figure 2 above for what it might look like pictorially and the table below for how it would be logged (note that $x_1$ and $x_2$ are censored intervals, meaning we had to cut them short due to them spilling out of the data processing window, while $t_1$, $t_2$ and $t_3$ are uncensored).}
\end{itemize}

\begin{figure}
  \includegraphics[width=0.8\linewidth]{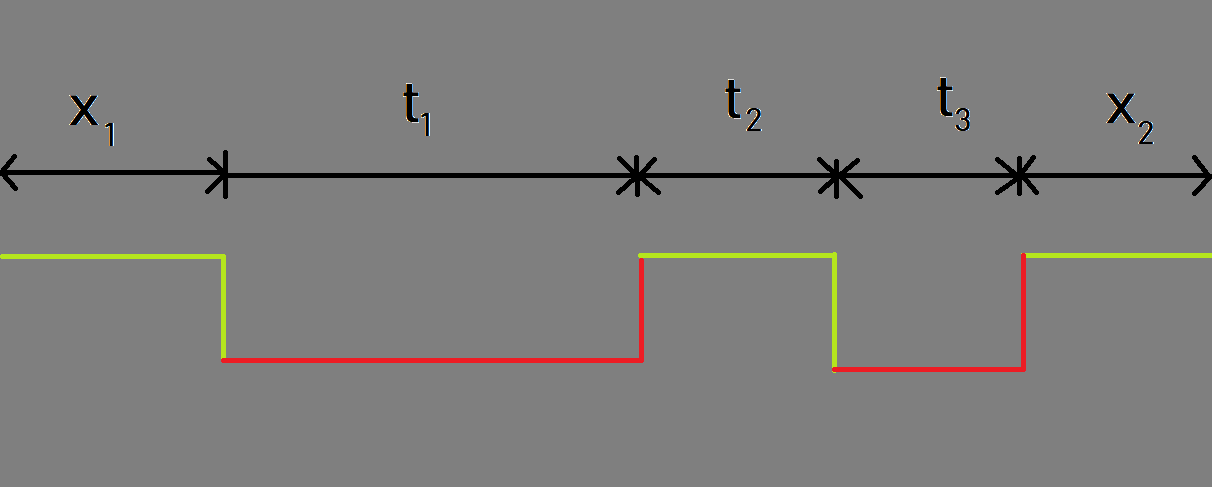}
  \caption{Toy state transition data}
  \label{topologies}
\end{figure}

\begin{center}
 \begin{tabular}{||c c||} 
 \hline
 State & Duration \\ [0.5ex] 
 \hline\hline
 UP & $x_1$ \\ 
 \hline
 DOWN & $t_1$\\
 \hline
 UP & $t_2$  \\
 \hline
 DOWN & $t_3$ \\
 \hline
 UP & $x_2$ \\ [1ex] 
 \hline
\end{tabular}
\end{center}

\subsection{Some more metrics}

In this section, we will define some new metrics for a general continuous-time, two-state process (UP and DOWN) like machine function. These will lead up to AIR. As the state transition data defined above hints, the process involves a machine  transitioning between two possible states (UP and DOWN). When it is UP, it stays in that state for a random amount of time before transitioning to a failed state. The average of this distribution can be described as Mean Time to Failure (MTTF). Likewise when the machine goes DOWN, it stays in that state for a random amount of time. The average amount of time time it stays in the DOWN state can be described as the Mean Time to Recovery (MTTR). 

The simplest assumption for such distributions is exponential, which has only one parameter (the rate). If we assume that the time to failure (TTF) is exponentially distributed with rate $\lambda$, we get an MTTF: $\frac{1}{\lambda}$. And if we assume that the time to recovery (TTR) is exponential with rate $\mu$, we get an MTTR: $\frac{1}{\mu}$. If we think of equation (2) and overall failure rate, we would first a time to failure (TTF), then a failure event followed by a time to recovery. Hence, on average there would be one failure every (MTTR+MTTF) units of time. So, the average rate would become:

$$\lambda = \frac{1}{\text{MTTR}+\text{MTTF}}$$

However, consider  that there would be two ways to bring this rate down. We either fail after very long times, increasing the MTTR or stay down for long periods of time, increasing the MTTF. But that doesn't make for a good KPI. If we're keeping customer VMs down for long periods, we don't want our KPI to become \textit{better}. It is thus better to simply get rid of the MTTR part and re-define failure rate as:

\begin{equation}\lambda = \frac{1}{\text{MTTF}}\tag{3}\end{equation}

Now, the only way to improve failure rate is to fail after ever-longer periods of time.

\subsection{Calculating MTTR and MTTF from heartbeat data}

Now, how to use the state transition data described earlier in this section to obtain the MTTR and MTTF (we only need MTTF for AIR but the methodology is very similar for the two). Let's start with the MTTF. If we knew the intervals between successive failures, we could simply average the lengths of these intervals. However, the problem with the state transition data described earlier in this section is that some of it is typically censored. For instance, if a component was UP for three days, it is unlikely to have one entry in the table with an UP duration of three days. This is because the data is typically processed for a day at a time. So, at the time we processed it at the start of the UP period, we only knew that the component had been UP for more than one day. We had no way of looking into the future and knowing that it would stay UP for three days. So, the data is censored (cut-short) and we just get an entry on the first day saying the component was UP for one day. Then on the second and third days, we get two more similar entries. 

So, the problem becomes that some of the intervals are un-censored (meaning we know for sure the component was UP for that duration) while other intervals logged in the data are censored (meaning we only know that the component was UP for a duration larger than the value that was logged).
The question becomes what to do with the rows in the data that correspond to censored observations. One way would be to try and look at the future to stitch together the various censored observations till they become un-censored. 

However, this quickly becomes messy and no matter how large an interval we take, we can never eliminate all the censored observations. Thankfully, there is a much simpler way to do this. It turns out, we can simply treat all censored and un-censored observations alike (a derivation is provided in appendix B).

If $n$ is the number of state transitions from the UP state to the DOWN state (or observed failures), $U_i$ are the duration's of the UP intervals from the state transition data of the component (some of them censored) and $D_j$ are the duration's of the DOWN components from the state transition data (again, some of them censored), we get the MTTF and MTTR respectively as:

\begin{equation}\frac{1}{\lambda} = \frac{\sum\limits_{i}U_i}{n}\tag{4}\end{equation}

\begin{equation}\frac{1}{\mu} = \frac{\sum\limits_{j}D_j}{n}\tag{5}\end{equation}

From equation (3), this means our failure rate becomes:

$$\lambda = \frac{n}{\sum\limits_{i}U_i}$$

And when the $U_i$'s are measured in 100-VM-years, the $\lambda$ becomes the AIR.

\chapter{Comparing AIR between groups}
Now, we have AIR as a KPI and know how to calculate it. We have an AIR number across the Azure fleet and all parts of the organization can work towards bringing it down. At the same time, there are multiple initiatives in Azure going on at any given time. All of them seek to improve some aspect of a very complex cloud system. How do we ensure none of them inadvertently cause the performance to regress (in truth, we can't completely eliminate this danger and regressions are indeed caused all the time; how do we detect as many of them as possible as early as possible)? The most likely times a system regresses is of course, the times when we change the system. Which loosely translates to new deployments of software components shipping features, fixing bugs and security vulnerabilities, etc. 

We then need to compare the number of failures that were being generated when the system was in a state without the deployment and the number of failures with the deployment. If the number of failures per unit time with the deployment is higher than the number of failures without, we can say that the deployment has regressed the failure rate of the system.

The problem is that we never know the actual rate of failures (per unit time). We can only estimate it from some finite amount of data we observed. And since the processes which generate these failures are noisy, stochastic processes, the failure rates we'll estimate will have some variance associated with them. In other words, if we were to repeat the process of observing some machines, measuring the up-time across machines and the number of failures and dividing them to get the rate again, we would get a slightly different number this time, even if the system is exactly the same.

The science of making decisions under this kind of uncertainty is called `hypothesis testing'. We'll go over the broad details in the next sub-section. If you want to dig more into hypothesis testing, refer to [2] for a comprehensive treatment and \href{https://towardsdatascience.com/hypothesis-testing-visualized-6f30b18fc78f?source=friends_link&sk=cd38bd44d242bb143cc184d6c2e6f0c1}{this blog} for an intuitive, visual introduction.

\section{Hypothesis testing}
The hypothesis testing problem can be formulated as follows: there are two populations. We want to tell if some statistical property (say, failure rate) of one of the populations is higher/lower/different than that of the other. The catch is that we can only estimate this statistical property for both groups and these estimates will be noisy. 

A logical thing to do would be to think of some number that quantifies the difference between the statistical properties. The most obvious is the literal difference of the properties, but another option could be taking the ratio. If the difference is zero, there is (by definition) no difference, but also if the ratio is one.

In practice, we only have estimates of our statistical property across the two groups and so, the difference, ratio or any other test-statistic we calculate on these estimates to test the difference will also be an estimate and hence involve some noise.

Depending on the distributions of the estimates of the statistical property, our test statistic will also have some distribution. It's natural to expect that as we collect more data, the variance in all these distributions should reduce.

\subsubsection{The null hypothesis and false positive rate}
Now, let's say there is no difference between the failure rates of the two populations. This is called the `null hypothesis' in Statistics. Also, let's say our test statistic is the difference between the two estimates of AIR for the two groups. If we frame our test so that we `reject the null hypothesis' and conclude there is a difference if we observe that the first group has a higher AIR estimate than that of the second group (or the difference in estimates which is our test statistic is greater than zero). It is natural to assume by symmetry that the test statistic will have an equal chance of being positive or negative. So, 50\% of the time, we will reject the null and conclude there is a difference between the AIR's of the two groups even though there is none. In other words, our test (which is set up to detect AIR differences) will erroneously fire about 50\% of the time. This becomes the false positive rate (FPR) of our test (denoted by $\alpha$). If this is too high for us, we can increase the bar and say we'll only fire the test if the difference exceeds some threshold greater than zero. Now, the probability that the test statistic will surpass this threshold by random chance will be lower than 50\% and so, our new test has a better false positive rate. If we were only concerned about false positives, we could extend this to its logical conclusion and make the threshold described above $\infty$. Then, our test statistic will never cross this threshold and so, if the null hypothesis is true, we will never incorrectly reject it. Notice however, that we achieved this feat by never predicting positives, making this a trivial and probably useless test because if the null hypothesis is \textit{not} true, the test will still never reject it. This brings us to our next section, but before we go there, note that the threshold for the test statistic isn't a static value, but is defined as the point where the probability that the test statistic will exceed that point is $p$, for some $p$ in our control (often set to 5\%). The reason for putting a threshold on $p$ instead of the test statistic itself is that our test starts to become receptive to ever smaller changes as we increase the data (and hence reduce the variance of the test statistic). In figure 2 below, Let's say there is indeed a difference between the AIR's for the two groups. This difference is given by the green dot. With the amount of data we collected, the pink line is to the right of the green point, so we will incorrectly conclude there is no difference (green point lies to the left of the threshold). However, as we increase the data in the two groups, the pink threshold where the area to the right is 5\% starts shifting to the left and we're eventually able to catch up with the green point.

\begin{figure}
  \includegraphics[width=0.8\linewidth]{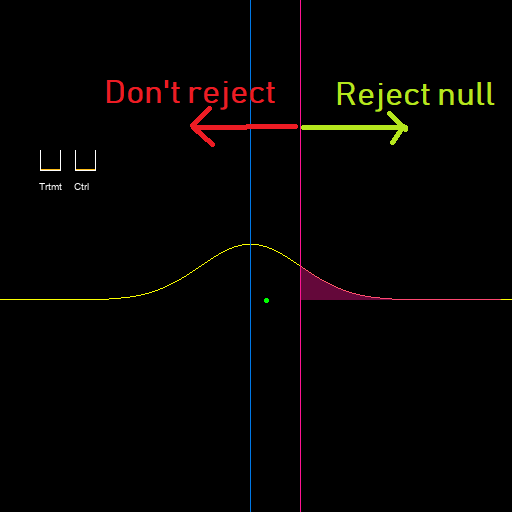}
  \caption{Hypothesis testing mechanism}
  \label{topologies}
\end{figure}

\begin{figure}
  \includegraphics[width=0.8\linewidth]{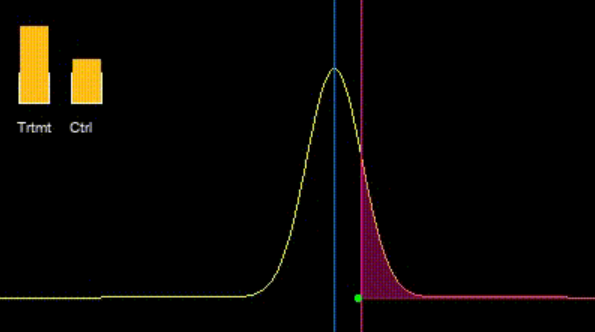}
  \caption{As we collect more data, we can catch smaller changes}
  \label{topologies}
\end{figure}

\subsubsection{The alternate hypothesis and false negative rate}
Now, we assume that the null hypothesis is \textit{not} true. So, the difference between the AIR numbers for the two groups is not zero. If it isn't zero, then what is it? This is actually a parameter we have to set and it's called the `effect size'. The smaller the effect size we set out to catch, the more data we need to catch it effectively. You can think of it as the resolution of our scale. If I only care about differences in length on the order of a centimeter, a regular measuring scale will suffice. But, if I want to be sensitive to differences of a nanometer, I'll have to spend a lot more money on a more sensitive instrument (the kind you would find in a Physics laboratory).

Once we pick our effect size, we know what the mean of our test statistic is (if the test statistic was the difference, the mean of the test statistic is the effect size itself). Like before, we can get the probability that the test will fail to reject the null hypothesis given the threshold (p-value) we're using as described in the previous section. This probability is called the false negative rate (FNR) and is denoted by $\beta$. We can see now why setting the threshold to $\infty$ (equivalent to p-value of 0) in the previous section was a bad idea. Even though we get a FPR of zero, we get a FNR of one (we never correctly identify a difference when it was there; which was the whole point of the test). And just like this was an easy way to get a perfect false positive rate, it's also easy to get a perfect false negative rate - just set the threshold to $-\infty$ (or p-value to one). So its clear that there is always a trade-off between the false positive and false negative rates. We can set the false positive rate as low as we like, but this always comes at the expense of a higher false negative rate. This trade-off is visualized in figure 4. The yellow distribution describes the test statistic under the null hypothesis (there is no difference in AIR's) and the purple distribution describes it under the alternate hypothesis (there is a difference and it equals the effect size we chose). The curve on the bottom-left of the figure shows the trade-off between these two metrics. By moving the central pink line to the right (lower the p-value); we get a better false positive rate but worse false negative rate. Moving it to the left (reducing the p-value) leads to a worse false positive rate, but better false negative rate.

\begin{figure}
  \includegraphics[width=0.8\linewidth]{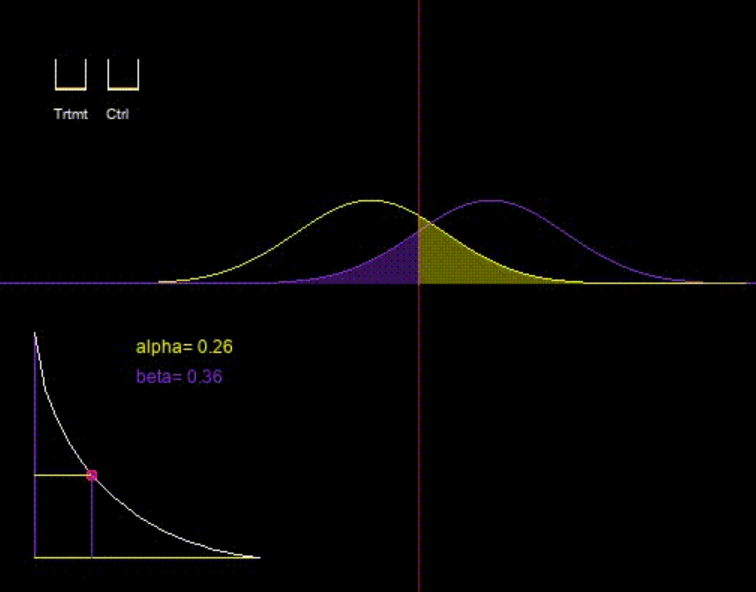}
  \caption{The false positive-false negative rate trade-off}
  \label{topologies}
\end{figure}

\subsubsection{Choosing the best test}
Let's say our current hypothesis test is giving us a false positive rate of 10\% and a false negative rate of 15\%. Let's also say that I want to reduce the false negative rate to 10\% without sacrificing on the false positive rate. Since there is always a trade-off between the false positive rate, should this be impossible? No - of course not. Even for the same hypothesis test, if we increase the amount of data we collect (which for AIR means observing the two groups for longer), we can improve our false negative rate without sacrificing on the false positive rate. But what if the amount of data we can collect is fixed? We can still improve our false negative rate without sacrificing our false positive rate by changing the hypothesis test we're using. Since 1-FNR is called the `power' of the test, a test that will have a better FNR given the same data and FPR is called a `more powerful test'. This begs the question - ``what is the most powerful test''? If possible, we would of course like to use that one. But we're getting somewhat ahead of ourselves here since that the answer to that question will depend on the stochastic process that our Azure VM-reboot data (or whatever our definition of interruptions is) follows.

\section{Count distributions for Azure reboots}

\subsection{The Poisson point process for modeling Azure VM reboots}
To measure AIR for Azure as a whole or any sub-population within it, we need to model the counts of interruptions, however they are defined (the most common being reboots). In other words, we need to find a distribution that models the counts of these interruptions within a given time interval. As discussed in section III-B, the stochastic process governing the availability state of a machine is actually a continuous-time, two-state process. This means that when machines go down, they stay down for a certain duration dubbed ``time to recovery'' (TTR). However, in Azure we generally have $\text{MTTF} >> \text{MTTR}$ (we probably wouldn't have a business otherwise). So, we can ignore the down duration's and think of them like point events.

The most basic and well-known such stochastic process for modeling the counts of events in time (which is also the building block of many other more complicated count processes) is the Poisson process (see chapter 5 of [1]). Here, we assume that the inter-arrival times of events follows an exponential distribution. This seems like a great fit for us since the exponential is the \textit{only} distribution that has a constant hazard rate with time. This is called the `memory-less' property (the process maintains no memory - the hazard rate remains the same regardless of what happened in the past).

The constant hazard rate of the exponential distribution also happens to be its only parameter ($\lambda$). And if such a process is observed for time, $t$, then the number of events is Poisson distributed with parameter $\lambda t$ (this parameter is the mean and also, incidentally, the variance of the Poisson distribution). See chapter 5 of [1] for details and derivations.

\subsubsection{Hypothesis test for Poisson point process}
There exists a uniformly most powerful (UMP) test for comparing two Poisson processes (see section 4.5 of [2]). This test (under certain assumptions) is guaranteed to produce the best false negative rate (power) given any false positive rate, effect size and amount of data. 

It basically becomes a test on the Binomial probability of successes and this test is implemented in programming languages like Python and R. 

We'd really like to use this test for comparing AIR numbers between two groups. But the question is, does Azure data follow the Poisson process? If it did, the number of reboots/ interruptions in any interval would be Poisson distributed. And the easiest way to spot a Poisson distribution is that the mean and variance are the same. For Azure reboots however, the variance of the counts in any given interval has been empirically observed to always be much higher than the mean (such distributions are called over-dispersed). And in the context of statistical hypothesis testing, the variance is a particularly important property, making this a potential deal-breaker.

So, since a Poisson process doesn't seem to have the observed property of variance being higher than mean, we will discuss in the next section some ways in which we might modify the Poisson point process so that we might get a higher variance than mean for the count of events in any given interval.

\subsection{Generalizations of the Poisson point process}
As mentioned in the previous section, the Poisson point process requires that the mean and variance of the count of events within any interval be the same. Since this assumption is too restrictive for Azure data, we'll look into two ways of generalizing it.

\subsubsection{Changing the inter-arrival distribution}
We mentioned earlier that in a Poisson process, the count of events in any interval of time is Poisson distributed while the inter-arrival times are exponentially distributed. We also mentioned that the exponential distribution is characterized by a constant failure rate. This means that it doesn't matter what happened upto a given instant of time; the rate of seeing new events is always the same (meaning the process is ``memoryless''). This might be restrictive since failures in a complex cloud system tend to often be clustered together (perhaps due to an outage, or just because multiple VMs are co-located on a node, other single points of failure like network switches, power units, etc.). So, given that a failure occurred the chances of other failures happening in general should increase. So, the hazard rate close to other failures should also be higher than it is when a lot of time has elapsed since the last one. This effect can be achieved if we simply pick a distribution for the inter-arrival times with decreasing hazard rate. When we see a failure, the hazard rate of the next failure spikes to its maximum before gradually tapering off. This makes the event arrivals more clustered (positively correlated) and causes more variance, which is the way Azure data behaves.

So, instead of the exponential distribution with its constant failure rate, we need to find another distribution with a decreasing failure rate. The Weibull distribution (see [3] for an entire book written on it) is a great candidate since it can model both monotonically increasing as well as decreasing hazard rates. So as we change our assumptions on hazard rate from constant to decreasing, this will also allow us to observe what would happen if we went the  other direction.

\begin{figure}
  \includegraphics[width=0.8\linewidth]{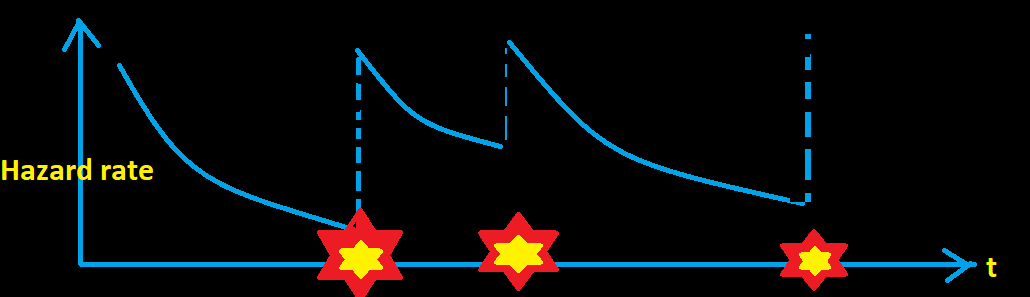}
  \caption{Decreasing hazard rate point process. An event arrival spikes the hazard rate, meaning other arrivals become more likely in the vicinity.}
  \label{topologies}
\end{figure}

\begin{figure}
  \includegraphics[width=0.8\linewidth]{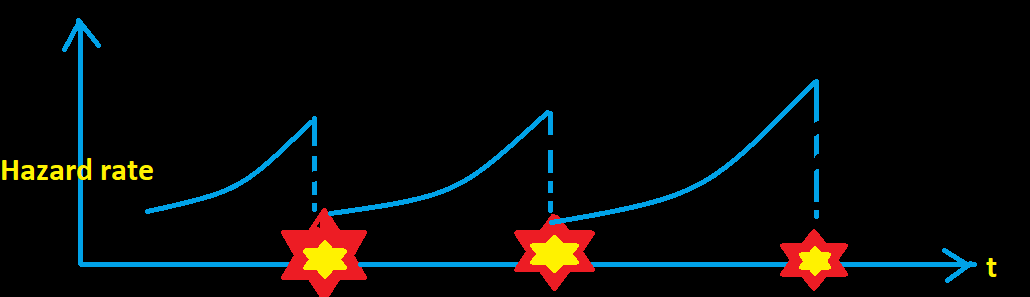}
  \caption{Increasing hazard rate point process. An event arrival causes the hazard rate to go back down, meaning other arrivals are pushed away.}
  \label{topologies}
\end{figure}

\begin{figure}
  \includegraphics[width=0.8\linewidth]{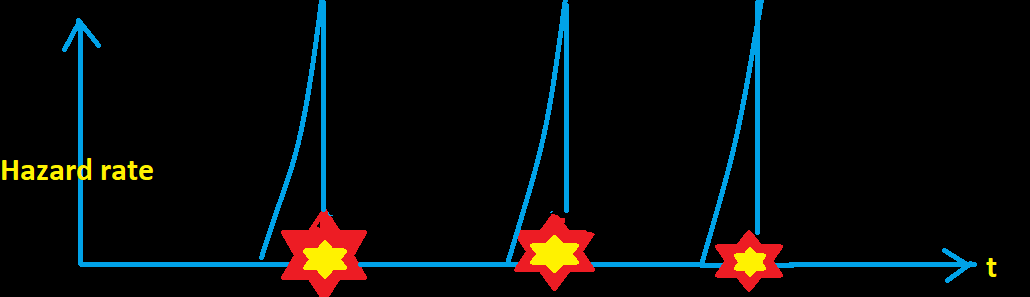}
  \caption{Extreme version of increasing hazard rate. This process represents events happening periodically.}
  \label{topologies}
\end{figure}

For a Poisson point process, the hazard rate is just a constant flat line and the mean and variance of the counts in any interval become equal.

Figure 2.4 shows a hazard rate profile where the inter-arrival distribution has a decreasing hazard rate. Each time an event occurs, the hazard rate spikes and other events become more likely. This imparts a positive correlation to the event arrivals, causing the count distribution to become over-dispersed (variance greater than mean). 

Figure 2.5 shows a hazard rate profile where the inter-arrival distribution has an increasing hazard rate. Each time an event occurs, the rate of seeing more events drops, pushing the event occurrences away from each other. This makes them negatively correlated, imparting a negative correlation and making the count distribution under-dispersed. 

Figure 2.6 takes this to the extreme and describes a process where the events happen every fixed interval of time. This means that the hazard rate spikes to $\infty$ at instants where the event deterministically happens. Since this is an extreme version of the case above, the variance becomes zero. And this makes sense since the point-process is not stochastic anymore.

All of these profiles can be modeled by the Weibull inter-arrival point process. Simulating from this count distribution is quite trivial. We just generate random Weibull inter-arrivals using the inverse transform technique as detailed in section 2.8.2 of [3] and count the number of arrivals within any time period. Unfortunately, we haven't been able to derive the probability mass function (PMF) or cumulative density function (CDF) of the resulting count distribution. Indeed, a closed form for these might not exist. But sampling from it is more than enough for our purposes here.

%TODO: Add some properties of this count distribution.

\subsubsection{The Compound Poisson Process}
As motivated in the previous section, there are multiple single points of failure within Azure that have the effect of clustering failures together, leading to a higher variance of event counts within any time interval than mean. The most obvious one is multiple VMs being co-hosted on a single physical machine (or node). If the node goes down, all the VMs will go down together. Now, the number of VMs on a node when it goes down will be a random variable itself and this motivates the Compound Poisson process (see section 5.4.2 of [1]). This is a simple extension of a Poisson process where every time we get a ``hit'' from the Poisson process, instead of having just one point-event, another random variable (say $Y$) dictates the number of point-events. For Azure, we can think of this as nodes downtime events following a Poisson process and every time a node goes down, another random variable dictates the number of VMs it brings down. 

Per equations (5.24) and (5.25) from [1], the mean and variance of such a point process will become (assuming the underlying Poisson process has a rate, $\lambda$):

$$E[X(t)] = \lambda t E[Y]$$
and,
$$ V[X(t)] = \lambda t E[Y]^2 $$

This allows for our variance to be much higher than the mean.

\section{Power of UMP Poisson test on Azure data}
Now that we've detailed two potential stochastic point processes that might model Azure data (with variance matching observations), let's try and measure how the hypothesis test we described in section 2.2.1 fares on them.

\subsection{UMP Poisson test on Compound Poisson process}
Let's see how the Poisson hypothesis test described in the previous section fares when applied to the Compound Poisson process. Chapter 4 of [2] proves that this is the uniformly most powerful test for comparing failure rates between two Poisson point process. 

Now, we're applying the test to a point process with a different distribution. Remember, the general construct of hypothesis testing involves applying a threshold to the p-value, which is the survival function of the test statistic. Let's call this threshold $\hat{\alpha}$. If we generate data from a process that is faithful to the assumptions of the test, then this $\hat{\alpha}$ is also the false positive rate of the test. If the data we generate is from a different process than the assumptions of the test, we can still apply it. However, any guarantees of the test being the ``most powerful'' are of course, void. In addition, since our assumption of the distribution of the test statistic is probably incorrect now, the $\hat{\alpha}$ we're using for the threshold is different from the actual false positive rate we'll observe. Let this rate be $\alpha$. It is easy to obtain the relationship between $\alpha$ and $\hat{\alpha}$ via simulation. We simply generate data from the null hypothesis (two sets of data from the same point-process). Then, we apply the test to these two data sets (we know the test shouldn't find a significant difference between them since we generated them from the same point process) and see what levels of $\hat{\alpha}$ would make the test erroneously fire (false positive). This can be achieved by maintaining an array of $\hat{\alpha}$'s ranging from $0$ to $1$ and seeing which of them are below the p-value in each simulation. We simply repeat this process many times and get the proportion of times each value of $\hat{\alpha}$ ``fired''. The resulting plot is shown in figure 2.7 for the cases where we apply the UMP Poisson test on Poisson data (blue line) and Compound Poisson data (orange line). First, observe that the blue line indicates that when we use the Poisson distribution, we always have $\alpha=\hat{\alpha}$ as expected. The orange line shows that when applying the test on the Compound Poisson process, the $\alpha$ first rapidly increases to $~0.3$, then gradually increases to $~0.6$ before shooting up to $1$ as $\hat{\alpha}$ increases from $0$ to $1$.

So, if we're targeting a certain false positive rate, $\alpha$ on a distribution like the Compound Poisson, we should ``look up'' the value of the $\hat{\alpha}$ to set use from the arrays generated by the simulation described above. One thing we need to be mindful of is the potential for this mapping to change depending on the sizes of the samples in the two groups we're comparing or the base rate, $\lambda$. Though we don't have a rigorous proof that it doesn't, we did do an experiment where we took various values of observation times and base rates in the two groups and observed the respective mappings. The results are plotted in figure 2.9. You can see that the mapping stays pretty stable, which is great news, since it means we can use the same mapping no matter what groups from the Compound Poisson we're comparing.

Now that we know how to achieve our target false positive rate, what about the false negative rate?

The process is very similar. For each possible $\hat{\alpha}$ in our array, we already have the actual $\alpha$'s from our simulation of the null hypothesis. Now, we pick some effect size and generate data from the alternate hypothesis. This is done by generating two sets of data once more. However while the first one is still from the original Compound Poisson process we used in the null hypothesis, the second one will have a higher value of $\lambda$ plugged into it (how much higher will be dictated by the effect size we're targeting). Now, we have samples generated from the alternate hypothesis. We know the test \textit{should} fire. We can plug the data from the two into our UMP for Poisson test and calculate the p-value. Then, like before, we get for each $\hat{\alpha}$ weather or not it would cause the test to fire (its value would be less than p-value) per simulation. This allows us to get the false negative rates per $\hat{\alpha}$ we could potentially set. And we already have the $\alpha$'s corresponding to each of them. So, we can now plot the $\alpha$-$\beta$ curve. When we do this for the UMP Poisson test applied to the Poisson and Compound Poisson distributions, we get a pleasant surprise. If you see figure 2.8 (notice just the blue and orange lines for now), you'll notice that the lines from the test applied to the Poisson and Compound Poisson point processes are more-or-less on top of each other. Surprisingly, the UMP Poisson test seems to be just as powerful for the Compound Poisson. The only consequence of having the wrong distributional assumption in this case seems to be that we can no longer close our eyes and expect to get the same $\alpha$ we set as a threshold in the test. For this, we need to `shift' it by using a lookup (one way to generate the lookup is to simulate data from the null hypothesis like described earlier in this section).

\begin{figure}
  \includegraphics[width=0.8\linewidth]{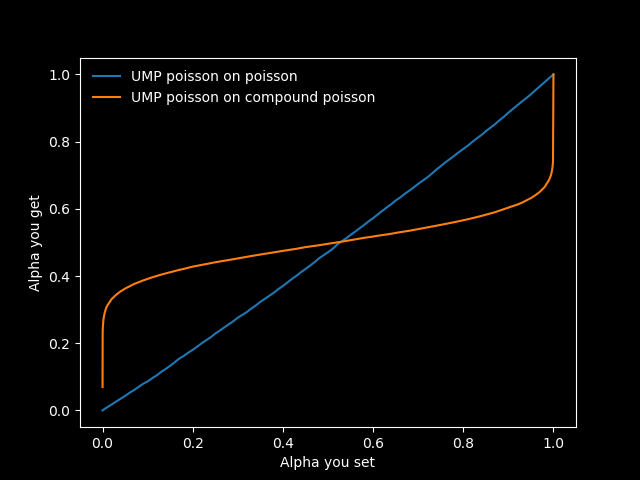}
  \caption{Relationship between the $\alpha$ you set and $\alpha$ you get}
  \label{topologies}
\end{figure}

\begin{figure}
  \includegraphics[width=0.8\linewidth]{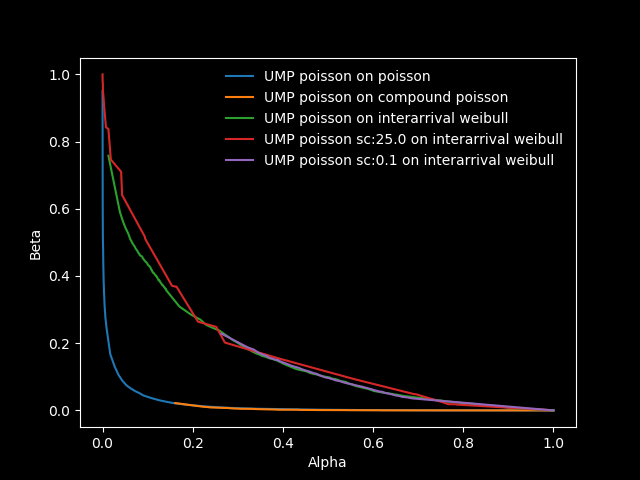}
  \caption{Alpha-Beta curves applied for the UMP Poisson test and some modifications on all point processes.}
  \label{topologies}
\end{figure}

\begin{figure}
  \includegraphics[width=0.8\linewidth]{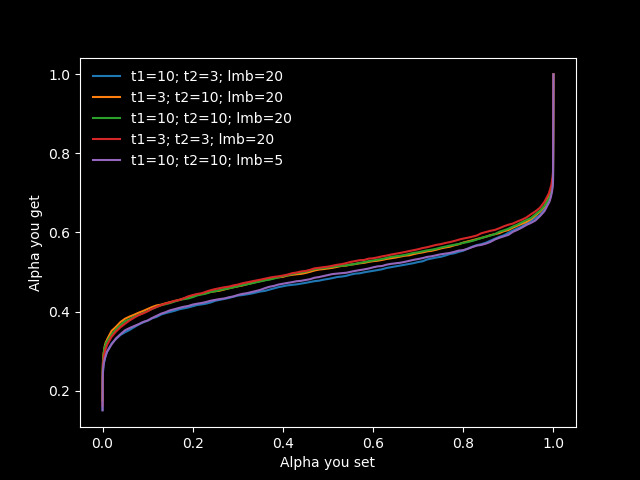}
  \caption{Does the mapping between $\hat{\alpha}$ and $\alpha$ change with the observation sizes of the two groups being compared when applying the UMP Poisson test to the Compound Poisson distribution? Thankfully, it doesn't.}
  \label{topologies}
\end{figure}

\subsection{UMP Poisson test on inter-arrival Weibull}
So, the UMP Poisson test doesn't seem to mind a Compound Poisson point process and generates just as much power a regular Poisson. Surely, it can't generate as much power for \textit{all} point processes. And indeed, when we apply it to the inter-arrival Weibull point process with $\kappa = 0.5$ (for the Weibull distribution), we get the green line in figure 2.8. You can see that this is significantly higher than the blue and orange lines from the Poisson and Compound Poisson point processes, meaning the false negative rate ($\beta$) is higher, making it less powerful on this point process.

Can we modify the test in some way to increase the power on this inter-arrival Weibull distribution? Here we describe an experiment that didn't work, but none-the-less provided insight into the workings of the test.

We note that the test simply takes the percentage contribution of the number of events from our first data set to the total events from both sets and compares it to the percent contribution of the time under observation for the first data set to the total and compares the two percentages. Meaning that if we scale the number of events in both groups by some scaling factor, we might get a slightly different kind of hypothesis test (the p-values certainly change if we do this).

This is exactly what we tried, both increasing and decreasing the scaling factor from $1$. The results are plotted in the red and purple lines of figure 2.8 (the red line corresponds to a scaling factor of $25$ and the purple one to $0.1$). Interestingly, these two lines are right on top of the green line. This means that by doing this scaling, we're changing the relationship between $\hat{\alpha}$ and $\alpha$ (the false positive rates we set and get respectively) but not between $\alpha$ and $\beta$, hence not changing the power of the test. This is reminiscent of the observation when we applied to UMP Poisson test to the Compound Poisson point process.

\section{Applications}
\subsection{Theoretically sound hypothesis test for AIR}
The result from the previous section that the UMP test for Poisson is just as powerful on the Compound Poisson is encouraging. This is because the compounding is probably the primary cause of the count distribution of failures within Azure being over-dispersed (higher variance than mean). If it were purely the hazard rate profile, then we would expect an increasing rather than decreasing hazard rate (failure becomes more likely as the machine runs longer). This should have resulted in an under-dispersed distribution (variance lower than mean) which is contrary to what we observe. If compounding is the primary cause of over-dispersion, it stands to reason that we should be in good standing using the UMP Poisson test, as long as we're mindful of the fact that the $\alpha$ we see isn't necessarily the $\alpha$ we get. The compounding process can be constructed carefully to match the mean and variance of the data we see within Azure and a ``lookup table'' that maps $\hat{\alpha}$ to $\alpha$ constructed. We can then set the $\hat{\alpha}$ corresponding to the desired $\alpha$ and be sure that the test we're using is well suited to the data, providing close to the best possible power. 

This exercise has been completed for Azure interruptions and the resulting function coded in Kusto (the data processing system of choice in Azure). It is being used extensively by teams to compare failure rates.

\subsection{Recommending time to wait}
The simulator described in section 2.3 was used there for two purposes:

\begin{itemize}
\item{Obtaining a mapping between $\alpha$ (desired FPR) and $\hat{\alpha}$ (actual FPR we should set).}
\item{Plotting the profile of $\alpha$ vs $\beta$.}
\end{itemize}

This was all assuming some effect size and with a certain amount of generated data (given the time intervals, $t_1$ and $t_2$ we'd be observing the two groups being compared for). Another related question is ``given the effect size and target false positive and false negative rates, how large should the intervals $t_1$ and $t_2$ be? This is really the opposite of asking ``given the false positive rate, effect size, $t_1$ and $t_2$, what false negative rate will be see?", so it's the same process from a different angle.

Of course, we'll get a whole curve in the $t_1$-$t_2$ space that satisfies the above requirements. If we have a constraint on one of them (for example, if we're comparing with a certain time-window from the past), we can run the simulation for various values of the other until we satisfy the requirements.

The question is, what hypothesis test and point-process should we plug into our simulator? The first question is easy, we should use the same test we intend to use to catch issues. For the second question, we discussed in the previous section that a Poisson distribution should be sufficient since the Compound Poisson and Poisson will involve equivalent statistical power with the UMP test.

Even if the primary cause of the over-dispersion in our data is the hazard rate profile, the time-to-wait from a Poisson assumption will always be lower than that from real data. So, we should wait \textit{atleast} the amount of time a Poisson distributional assumption recommends for reaching certain target false positive and false negative rates for over-dispersed data in any case. 

An online simulator was prepared for this purpose and it is being used to determine the time to wait between various stages of a deployment.

%\begin{appendices}
\appendix

\chapter{Derivation of hazard rate function}
% the \\ insures the section title is centered below the phrase: AppendixA

Let's say that a process produces some events of interest (like motor accidents, machine failures, etc.). The time between successive occurrences of such events (inter-arrival time) is a random variable. Let's call it $T$. The probability density function (PDF) of this random variable is denoted by $f_T(t)$. By definition, the probability that we will see an event between some interval $(t, t+\delta t)$ is given by:

$$P(T \in (t,t+\delta t)) = f_T(t) \delta t$$

A quantity that is more useful in many contexts than the PDF is the hazard rate.

Let's say someone has had successful cancer treatment. The sad thing about cancer is that it is never completely cured and there is always a chance the body will relapse. Given this, a cancer patient might wonder: "it's been 1 year since my treatment and I haven't relapsed yet. Given I didn't relapse until now, what is the chance I'll relapse in the next month?". We can even remove the arbitrary 'month' interval in this statement and simply ask how many events do I expect to see per unit time. This becomes a 'rate' which is similar conceptually to velocity. Just as we have instantaneous velocity, we have instantaneous rate. We can express this notion mathematically as:

$$P(T \in (t,t+\delta t)\:| \: T > t) = \frac{P(T \in (t,t+\delta t) \,  \& \, T>t)}{P(T>t)}$$

$$= \frac{P(T \in (t,t+\delta t))}{P(T>t)}$$
By definition of the probability density function, $f_T(t)$ this becomes:

$$ = \frac{f_T(t) \delta t}{P(T>t)} $$

As $\delta t$ becomes small, the probability that more than one event will occur in that interval becomes negligible. So, there will be either $0$ or $1$ events in this interval when it is sufficiently small, effectively making the event a coin toss (a.k.a. a Bernoulli random variable). So, the probability calculated above is also the expected number of events in the small interval. Then, if the number of events per unit time is defined as the hazard rate function, $h_T(t)$; the number of events in a small interval proceeding $t$ will become: $h_T(t)\delta t$. Equating the two expressions we get:

$$h_T(t)\delta t = \frac{f_T(t)}{P(T>t)} \delta t$$

simplifying,

$$h_T(t) = \frac{f_T(t)}{P(T>t)} = \frac{f_T(t)}{S_T(t)}$$

Now, this rate is a function of time. Which means that in any given large enough interval of time, it will take on different values. What if we wanted to approximate it with a single number (say $\lambda$) for a given interval? By definition of averages, we would have:

$$\int_{t_1}^{t_2} h_T(t) dt = \int_{t_1}^{t_2} \lambda dt = \lambda(t_2-t_1)$$

Also, let's say that the number of events observed in the interval $(t_1,t_2)$ is given by the random variable, $N$. Then the expected value of $N$ is given by (by definition of $h_T$):

$$E(N) = \int_{t_1}^{t_2} h_T(t) dt = \lambda (t_2-t_1)$$

So we get:

\begin{equation}\lambda = \frac{E(N)}{t_2-t_1} \tag{2}\end{equation}

If we observe the process for a certain interval of time and count the number of events, $n$ within said interval then $n$ is an unbiased unbiased estimator for $E(N)$ and so the estimator for $\lambda$ becomes:

$$\hat{\lambda} = \frac{n}{t_2-t_1} = \frac{n}{\bigtriangleup t}$$

\chapter{Using the exponential distribution to estimate AIR}

Not only is the exponential distribution the simplest possible distribution for modeling the time until some event, it is also the only distribution that has a constant failure rate. This makes it a natural choice for estimating a single failure rate. Here, we will use it to obtain an unbiased estimator for the failure rate when some of our data is censored and some is un-censored. 

First, two quick properties of the exponential distribution. The probability density function is given by:

$$f_T(t) = \lambda e^{-\lambda t}$$

And the survival function is given by:

$$F_T(t) = \int\limits_t^\infty f_T(t) dt = e^{-\lambda t}$$

Using these, the likelihood function of data with $t_i$ un-censored data and $x_j$ censored data points becomes:

$$L(\lambda) = \prod\limits_{i=1}^n \lambda e^{-\lambda t_i} \prod\limits_{j=1}^m e^{-\lambda x_j}$$

Taking logs on both sides, we get the log likelihood function.

$$ll(\lambda) = \sum\limits_{i=1}^n (\log \lambda -\lambda t_i) - \sum\limits_{j=1}^m \lambda x_j$$

To get the value of the parameter $\lambda$ that minimizes the log-likelihood function, we take derivative with respect to $\lambda$ and set it to zero.

$$\frac{\partial ll(\lambda)}{\partial \lambda} = \frac{n}{\lambda} - \left(\sum\limits_i t_i + \sum\limits_j x_j\right)$$

$$\frac{\partial ll(\lambda)}{\partial \lambda} = 0$$

This gives us the MTTF:
$$\frac 1 \lambda = \frac{\sum t_i +\sum x_j}{n}$$

Hence, we can simply sum all the UP times (censored and uncensored) and divide by the number of downtime events to get the MTTF. A similar result holds for MTTR, replacing UP times with DOWN times. 

The AIR estimate then becomes: $\frac{1}{\text{MTBF}}$, which is just the total number of failures divided by the total UP time.

%\end{appendices}

\end{document}